\documentclass[12pt]{iopart}
\usepackage{graphicx}
\usepackage{subfigure}
\expandafter\let\csname equation*\endcsname\relax    %宏包iopart和amsmath冲突
\expandafter\let\csname endequation*\endcsname\relax %宏包iopart和amsmath冲突
\usepackage{amsmath}
\usepackage{physics}
\usepackage{iopams}
\usepackage{tabularx}
\usepackage{enumerate}
\usepackage[colorlinks,linkcolor=blue,citecolor=blue,urlcolor=blue]{hyperref}

\begin{document}

\title{Generic non-Hermitian mobility edges in a class of duality-breaking quasicrystals}
% Rewrite this paragraph in an academic language: [PARAGRAPH]
% Paraphrase the text using more academic and scientific language. Use a neutral tone and avoid repetitions of words and phrases:[PARAGRAPH]
% Correct the grammar: [PARAGRAPH]
% Please proofread and polish the passage from an academic angle and highlight the modifications:
% Here is a paragraph from an academic paper. Please polish it to conform to academic standards, enhancing its spelling, grammar, clarity, conciseness, and overall readability. If needed, sentences should be rewritten. Additionally, list all modifications and their reasons in a Markdown table：
% Please rewrite the following content in nature journal style:

\author{Xiang-Ping Jiang $^{1}$, Mingdi Xu $^{2}$ and Lei Pan $^{2,*}$}

\address{$^1$ Zhejiang Lab, Hangzhou 311121, China}
\address{$^2$ School of Physics, Nankai University, Tianjin 300071, China}
\address{$^*$ Author to whom any correspondence should be addressed.}

% \ead{2015iopjxp@gmail.com}
% \ead{2120230195@gmail.nankai.edu.cn}
\ead{panlei@nankai.edu.cn}

\vspace{10pt}
\begin{indented}
\item[]
\end{indented}

\begin{abstract}
We provide approximate solutions for the mobility edge (ME) that demarcates localized and extended states within a specific class of one-dimensional non-Hermitian (NH) quasicrystals. These NH quasicrystals exhibit a combination of nonreciprocal hopping terms and complex quasiperiodic on-site potentials. Our analytical approach is substantiated by rigorous numerical calculations, demonstrating significant accuracy. Furthermore, our ansatz closely agrees with the established limiting cases of the NH Aubry-Andr{\'e}-Harper (AAH) and Ganeshan-Pixley-Das Sarma (GPD) models, which have exact results, thereby enhancing its credibility. Additionally, we have examined their dynamic properties and discovered distinct behaviors in different regimes. Our research provides a practical methodology for estimating the position of MEs in a category of NH quasicrystals that break duality.
\end{abstract}

% \vspace{2pc}
\noindent{\it Keywords}:  Anderson transition, mobility edge, non-Hermitian quasicrystal, non-equilibrium dynamics
%
% Uncomment for Submitted to journal-title message
\maketitle

%%%%%%%%%%%%%%%%%%%%%%%%%%%%%%%%%%%%%%%%%%%%%%%%%%%%%%%%%%%%%%%%%%%%%%%%
%%%%%%%%%%%%%%%%%%%%%%%%%%%%%%%%%%%%%%%%%%%%%%%%%%%%%%%%%%%%%%%%%%%%%%%%
\section{Introduction}
Anderson transition is a crucial concept in condensed matter physics that describes the suppression of coherent metallic transport due to quenched disorder, leading to a metal-insulator transition~\cite{anderson1958absence}. This phenomenon arises when the extended states of a clean system become exponentially localized by the disorder. The quenched disorder can induce an extended-localized transition at a critical energy called the mobility edge (ME)~\cite{thouless1974electrons,lee1985disordered,kramer1993localization,evers2008anderson}, marking the boundary between localized and extended states in three dimensions. However, this transition does not occur in one and two-dimensional disordered systems, as confirmed by the well-established one-parameter scaling theory~\cite{abrahams1979scaling}. Despite this, localization transition can emerge in one-dimensional (1D) quasiperiodic systems, such as those exhibiting incommensurability with the underlying lattice. A prime example of a 1D quasiperiodic lattice is the Aubry-Andr{\'e}-Harper (AAH) model~\cite{harper1955single,aubry1980analyticity}, which exhibits a localization transition dependent on quasiperiodic potential. The model exhibits exact self-duality in the real and momentum space and lacks MEs in the single-particle spectrum. If this fine-tuned self-duality is destroyed, the energy-dependent MEs may appear in generalized AAH models~\cite{sarma1988mobility,biddle2009localization,biddle2010predicted,ganeshan2015nearest,danieli2015flat,deng2019one,li2017mobility,wang2020one,wang2020realization,wang2021duality,liu2022anomalous,wang2022quantum,gonccalves2022hidden,gonccalves2023critical,gonccalves2023renormalization,vu2023generic,wang2023engineering,qi2023multiple,zhou2023exact,borgnia2022rational,borgnia2023localization}. Thus, these predicted MEs have sparked significant interest and have been experimentally observed~\cite{roati2008anderson,luschen2018single,an2018engineering,an2021interactions,lin2022topological,wang2022observation,li2023observation}.

Recently, studies of the non-Hermitian effect on Anderson transitions and MEs have attracted significant interest in disordered and quasiperiodic systems~\cite{zeng2017anderson,longhi2019topological,longhi2019metal,jiang2019interplay,zeng2020topological,liu2020non,liu2020generalized,tzortzakakis2020non,huang2020anderson,schiffer2021anderson,tang2021localization,liu2021localization,liu2021exact,liu2021exact1,cai2022localization,jiang2021mobility,jiang2021non,wu2021non,cai2022equivalence,sarkar2022interplay,zeng2022real,jiang2023general,qi2023localization,padhan2024complete,acharya2024localization,jiang2024localization,jiang2024exact,jiang2024exact1}. In general, the absence of Hermiticity is typically achieved through the introduction of nonreciprocal hopping processes or gain and loss terms, which may give rise to exotic phenomena that lack Hermitian counterparts. Examples of such phenomena include exceptional points, parity-time ($\mathcal{PT}$) symmetry broken, and non-Hermitian skin effect~\cite{yao2018edge,gong2018topological,lee2019anatomy,okuma2020topological,zhang2020correspondence,borgnia2020non,song2019non,yi2020non,guo2021exact,longhi2021phase,zeng2022real1,longhi2022non,peng2022manipulating,mao2023non,lin2023topological,mao2024liouvillian}. The interplay between non-Hermiticity and quasiperiodicity has been studied in various extensions of AAH models, in which the introduction of nonreciprocal hopping or complex on-site quasiperiodic potential~\cite{zeng2020winding,zhou2022driving,xia2022exact,xu2022exact,yuce2022coexistence,xu2021non,zhou2022topological,peng2023power}.  There exist several well-known examples where analytical solutions for MEs have been derived. One such example is the NH Ganeshan-Pixley-Das Sarma (GPD) model~\cite{ganeshan2015nearest,wang2021duality,liu2021exact}, where the kinetic-energy hopping term in the AAH model exhibits a long-range spatial form but decays exponentially. By employing a prior dual transformation, an analytic ME can be constructed within this model. Another example is the model presented in  Refs.~\cite{wang2020one,liu2021exact1}, where the quasiperiodic potential is mosaic, and the exact ME is derived through computing the Lyapunov exponents from Avila’s global theory~\cite{avila2015global,avila2017sharp}. These models represent distinct generic classes of generalized AAH models, wherein either the hopping kinetic energy or the incommensurate potential term is modified. However, the inverse problem—locating the MEs given a general duality-breaking NH quasiperiodic Hamiltonian—has not been thoroughly investigated.

In the present paper, we address the above questions and present a feasible theoretical method to estimate the MEs separating localized and extended states for a class of duality-breaking quasicrystals with both nonreciprocal hopping terms and complex on-site potentials. Furthermore, we have validated our general ME results through numerical simulations that compare with our theoretical ansatz. While our ansatz is not exact, it closely approximates the MEs, with discrepancies of only a few percentage points, even when the ME exhibits complex and non-monotonic structures in the parameter space of the quasiperiodic potential strength. Our findings highlight the rich variety of localization behavior observed in quasiperiodic systems compared to simpler random disorder-induced Anderson localization. This theoretical result has the potential for direct experimental verification, thereby advancing our understanding of localization behaviors in 1D  quasiperiodic systems.

The remainder of this paper is structured as follows. Firstly, we introduce a class of NH dual-breaking quasiperiodic models in Sec.~\ref{section2}. Subsequently, we present analytical approximations to estimate the MEs in Sec.~\ref{section3}. The numerical evidence of the MEs is demonstrated in Sec.~\ref{section4}. In Sec.~\ref{section5}, we investigate the dynamical properties of these quasicrystals holding ME. Lastly, a summary is provided in Sec.~\ref{section6}. 

%%%%%%%%%%%%%%%%%%%%%%%%%%%%%%%%%%%%%%%%%%%%%%%%%%%%%%%%%%%%%%%%%%%%%%%%
%%%%%%%%%%%%%%%%%%%%%%%%%%%%%%%%%%%%%%%%%%%%%%%%%%%%%%%%%%%%%%%%%%%%%%%%
\section{The Model Hamiltonian}\label{section2}
We consider a tight-binding NH quasiperiodic model, whose Hamiltonian is described by:
\begin{equation}\label{eq1}
    H=\sum_{j=1}^L(t_{L}c_{j}^\dagger c_{j+1} + t_{R}c_{j+1}^\dagger c_{j}) + \sum_{j=1}^L W(2\pi\beta j + \theta)c_j^\dagger c_j,
\end{equation}
where $c^{\dagger}_{j}(c_{j})$ denotes the creation (annihilation) operator for spinless fermions at site $j$, and $t_{L(R)}= te^{\pm g}$ represents the left (right) hopping amplitude, with $g$ quantifying non-reciprocity. The phase factor $\theta=\phi +ih $ is complex and the on-site potential $W(x)=W(x+2\pi)$ is periodic but incommensurate with the lattice for an irrational number  $\beta=(\sqrt{5}-1)/2$. When the potential $W(2\pi \beta j + \theta) = V\cos(2\pi \beta j + \theta)$ and the NH parameters $g=h=0$,  the Hamiltonian (\ref{eq1}) reduced to the well-known exactly solvable AAH model. The AAH model has an energy-independent localization transition and is self-dual at the critical point $V=2$, there are no ME in the whole spectrum. 

In this work, we are interested in a class of duality-breaking quasicrystals with complex potentials, which is given by a series of cosine functions with arbitrary coefficients:
\begin{equation}\label{eq2}
	W(2\pi\beta j + \theta) = V\cos(2\pi\beta j + \theta)\left[1+\sum_{m=1}^{M}\alpha_m \cos^{m}(2\pi\beta j + \theta)\right],
\end{equation}
where $V$  is the strength of the quasiperiodic potential,  $ M\geq 1 $ is an integer, and the coefficients ${\alpha_m}$ are real numbers. Without loss of generality, we take $t=1$ as units of hopping energy and choose the periodic boundary condition (PBC) for the following calculation.

%%%%%%%%%%%%%%%%%%%%%%%%%%%%%%%%%%%%%%%%%%%%%%%%%%%%%%%%%%%%%%%%%%%%%%%%
%%%%%%%%%%%%%%%%%%%%%%%%%%%%%%%%%%%%%%%%%%%%%%%%%%%%%%%%%%%%%%%%%%%%%%%%
\section{Mobility edge ansatz}\label{section3}
We consider a general periodic potential $W(2\pi\beta j + \theta)$ that is $W(x)=W(x+2\pi)$ under $2\pi$ translation. When the lattice $j\in \mathbb {Z}$ and the parameter $\beta$ is irrational, the on-site potential never repeats itself, making the lattice system quasiperiodic. This complexity poses a challenge in directly solving the model since the Bloch theory cannot be applied when $\beta$ is incommensurable. One trick approach is to approximate $\beta$ as $\beta\approx n_2/n_1$ (with $\beta<1$  implying $n_2<n_1$) and reduce the system size to  $n_2$ or less. In this case, the potential becomes periodic under $n_1-$site translation, enabling the use of the Bloch wave ansatz. However, this period is larger than the system size, making it physically aperiodic. This analytical technique, used by Aubry and Andr{\'e}, facilitates the derivation of the well-known duality in the AAH model, where self-dual points mark the transition between localized and extended states. 

A similar duality called hidden duality~\cite{gonccalves2022hidden,gonccalves2023critical, gonccalves2023renormalization,vu2023generic} can also be created in these quasiperiodic models using rational approximation. By assuming  $\beta\approx n_2/n_1$, one can consider a lattice with an enlarged unit cell of $n_1$ physical sites. This lattice can be solved by rescaling Bloch momentum $\kappa=k/n_1 \in [-\pi,\pi]$. Due to translational invariance under shifting $n_1$ sites, the phase $\phi$ can also rescaled into $\varphi=\phi/n_1\in [-\pi,\pi]$. Then, the Fermi surface in the 2D $(\kappa,\varphi)$ phase space can be solved by solving the following Hamiltonian matrix:
\begin{equation}
	H(\kappa,\varphi;g,h)=
	\left(
	\begin{array}{cccccc}
		W(\frac{\varphi}{n_1}+ih) & 1 & 0 & 0 & \cdots & e^{i\kappa+ig} \\
		1 & W(\frac{2\pi n_2+\varphi}{n_1}+ih) & 1 & 0 & \cdots & 0 \\
		% 0 & 1 & W(\frac{2\pi n_2+ \varphi}{n_1}+ih) & 1 &\cdots & 0 \\
		\vdots & \vdots & \vdots & \vdots  & \ddots & \vdots\\
		e^{-i\kappa-ig} & \cdots & 0 & 0 & 1 & W(\frac{2\pi n_2(n_1-1)+ \varphi}{n_1}+ih)
	\end{array}
	\right),
\end{equation}
where the quasiperiodic potential $W(\phi+ ih)$ with $\beta$ replaced by $n_2/n_1$. When we consider the NH parameters $g=h$, it can be hypothesized that in the limit $n_1, n_2\to\infty$ and at the hidden-dual point $(E, V)$, the Fermi surface becomes invariant under the exchange $[\cos(\kappa+ig) \leftrightarrow \cos(\varphi+ih)]$. While this duality is not held for the finite $n_1$, numerically obtaining a best-fit point that demonstrates strong invariance under $\varphi\leftrightarrow\kappa$ exchange converges to the true self-dual point as $n_1$ increases.

For analytical testing of the commensurate approximation in the NH AAH and GPD quasiperiodic models, we start with the simplest case of $n_1 = 1$. Then consider the NH AAH model $W(\varphi+ ih) = V\cos(\varphi + ih)$, thus we obtain the resulting energy expression
\begin{equation}
		E(\kappa,\varphi;g,h) = 2\cos(\kappa + ig) + V\cos(\varphi + ih), 
\end{equation}
which is self-dual for $V=\pm 2$. Similarly, for the NH GPD model $W(\varphi+ ih) = V\cos(\varphi + ih)/[1-\alpha\cos(\varphi + ih)]$,  we can also derive an expression
\begin{equation}
	\begin{split}
			&E(\kappa,\varphi;g,h) = 2\cos(\kappa + ig) + V\cos(\varphi + ih)/[1-\alpha\cos(\varphi + ih)]  \\
			& \Leftrightarrow (E\alpha+V)\cos(\varphi + ih) + 2\cos(\kappa + ig) - 2\alpha\cos(\varphi + ih)\cos(\kappa + ig) - E = 0,
	\end{split}
\end{equation}
which is self-dual at $V = \pm 2 - \alpha E$.
In the preceding examples of analytical duality, the potential was characterized by either a single cosine harmonic (the NH AAH model) or an exponential series (the NH GPD model). In this paper, we focus on a more generic scenario, where a series of cosines with arbitrary coefficients represent the potential:
\begin{equation}\label{equation6}
	W(\varphi+ ih) = V\cos(\varphi+ ih)\left[1+\sum_{m=1}^{M}\alpha_m \cos^m(\varphi+ ih) \right].
\end{equation}
For the generic set of $\{\alpha_m\}$,  the Fermi surface at an $n_1-$rational approximant is defined as the solution of the following equation:
\begin{equation}\label{equation7}
	P_{n_1}[W(\varphi+ ih);E,V] + 2\cos(\kappa+ig) = 0.
\end{equation}
This function, denoted as $P_{n_1}[W(\varphi+ ih);E,V]$, is an $n_1-$-order polynomial in the variable $W(\varphi+ ih)$, with the energy $E$ and the potential strength $V$ as parameters. However, when we consider the specific case where $n_1$ is finite, finding an exact dual point becomes a challenging task due to the distinct powers of the cosine functions involved. The cosine functions $\cos(\kappa+ig)$ and $\cos(\varphi+ ih)$ behave differently, making it difficult to find a balance where both sides remain invariant under the transformation $\varphi\leftrightarrow\kappa$. As $n_1$ approaches infinity, the polynomial function begins to simplify, eventually transforming into a different function (often through a Taylor series expansion), enabling the attainment of exact duality.  This simplification allows us to achieve an exact duality that was previously unattainable. At $n_1=1$, from the Eqs.~(\ref{equation6}) and (\ref{equation7}), we arrive at a simplified energy equation:
\begin{equation}\label{equation8}
	\begin{split}
		E(\kappa,\varphi;g,h)= 2\cos(\kappa+ig)  + V\cos(\varphi+ih)\left[1+\sum_{m=1}^M \alpha_m \cos^m(\varphi+ih)\right].
	\end{split}
\end{equation}
The path to mathematical harmony seemed to require the entire Fermi surface to remain unchanged under the exchange of $\varphi\leftrightarrow\kappa$ when the parameters $g=h$. Yet, this was an unattainable ideal. Instead, we focused on the real part of the energy and the cosine functions, seeking solace in the real part of the energy and the conditions $\cosh(g) \cos(\kappa)=0$ or $\cosh(h)\cos(\varphi)=0$. For instance, if $\cos(\varphi)=0$, then $\cos(\kappa)=E_{R}/2$, allowing us to map this point to to $\cos(\varphi) = E_{R}/2$ and $\cos(\kappa) = 0$, yielding
\begin{equation}\label{equation9}
	E_{R} = \frac{E_{R}V}{2}\left[1 + \sum_{m=1}^M \alpha_m \left(\frac{E_{R}}{2}\right)^m \right].
\end{equation}
Thus, the analytical approximated ME can be expressed
\begin{equation}\label{equation10}
	V = 2\left[1+\sum_{m=1}^M \alpha_m \left(\frac{E_{R}}{2}\right)^m\right]^{-1}.
\end{equation}
This approximation is incredibly useful because it allows us to quickly and easily understand the behavior of the system without needing to perform complex calculations. However, it's important to remember that this method is only valid for certain conditions and parameter values, yields that the exchange $\cos(\varphi) \leftrightarrow \cos(\kappa)$ only uses for the parameters $g=h$ and is suitable for the real part of the energy $E_{R}$. There is also another ME for $V<0$ when exchanging $\cos(\varphi)\leftrightarrow - \cos(\kappa)$, but for conciseness, we only focus on the $V>0$ side. The quest for a perfect match extended beyond the confines of our current approximation, inviting us to explore higher-order approximations of $\beta$ corresponding to larger $n_1$, which will most likely improve the ME prediction. Yet, as we delved deeper, the complexity of the terrain increased, with multiple powers of the $\cos\kappa$ and $\cos\varphi$, threatening to obscure the path. While such an approach might refine our predictions, the resulting complexity could render the theory less accessible and practical for use. Subsequently, it is observed that for the case where $n_1=1$, the analytically derived ME (\ref{equation10}) prediction exhibits a commendable level of accuracy.

\begin{figure*}[t]
    \centering
    \includegraphics[width=0.98\textwidth]{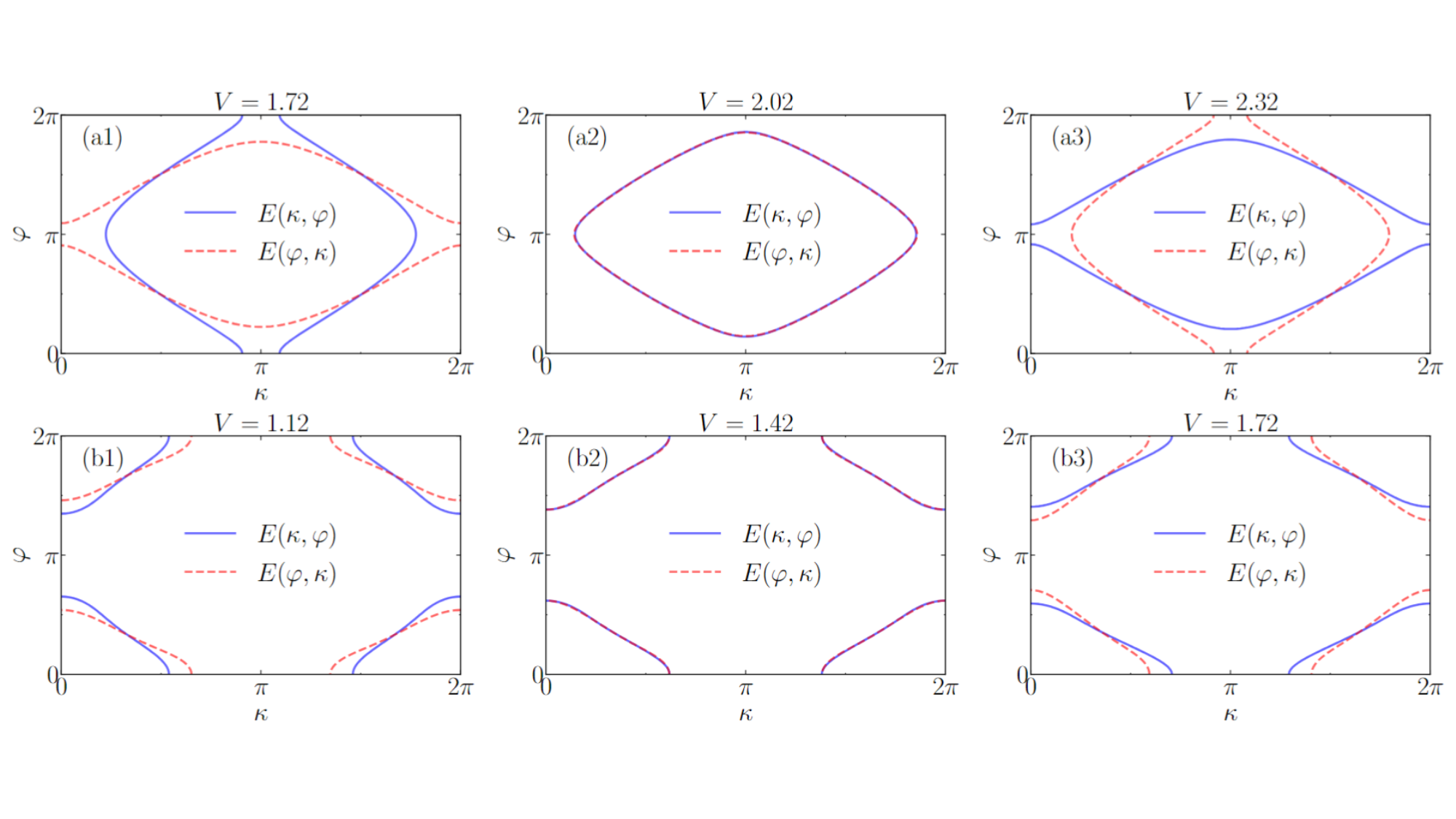}
    \vspace{-0.3in}
    \caption{Fermi surfaces at $E = 0$, $\alpha_1=0.1$ (a) and $E = 1.75$, $\alpha_1=0.2$, $\alpha_2=0.3$ (b) with the $\varphi \leftrightarrow \kappa$ exchange. The Fermi surface is invariant under the $\varphi \leftrightarrow \kappa$ exchange for (a2) and (b2) at approximately $V \simeq 2.02$ and $V \simeq 1.42$, respectively.}
    \label{fig1}
\end{figure*}

We now benchmark our ansatz~\ref{equation10} against the numerical self-duality. For this purpose, we set $\beta=(\sqrt{5}-1)/2$ for simplicity so that $\beta$ can be progressively approximated by $F_{j-1}/F_{j}$ where $F_j$ is a Fibonacci number and we set the lattice size $L=F_j$ and numerical calculation under the PBC. We first check the numerical invariance of the Fermi surface under $\varphi\leftrightarrow\kappa$ exchange at $g=h$ and $n_1=1$ approximant. For $\alpha_1=0.1$ and $E=0$, Eq.~(\ref{equation10}) predicts $V_c=2.0$. As shown in Fig.~\ref{fig1}(a2) for the value $V=2.02$, this value produces a visibly invariant Fermi surface as compared to the adjacent values $V=1.72$ and $V=2.32$ of Figs.~\ref{fig1}(a1) and (a3). The value of this invariant Fermi surface is very close to the analytically predicted value of $V_c=2.0$ and an error of $1\%$. For $\alpha_1=0.2, \alpha_2=0.3$ and $E=1.75$, Eq.~(\ref{equation10}) predicts $V_c=1.424$, as shown in Fig.~\ref{fig1}(b2), for the value $V=1.420$, this value produces a visibly invariant Fermi surface as compared to the adjacent values $V=1.12$ and $V=1.72$ of Figs.~\ref{fig1}(b1) and (b3). The value of this invariant Fermi surface is very close to the analytically predicted value of $V_c=1.424$ and an error of $0.3\%$. These results are remarkable, given that these cases have no other theoretical prediction or analytical solution before. Note that here we only give a few energies and give a few corresponding examples to draw the critical $V$ that satisfies the exchange invariance $\varphi\leftrightarrow\kappa$ and compare it with the analytical $V_c$ that predicted by Eq.~(\ref{equation10}). The critical $V_c$ is not the same for different energies so the error will be slightly different.

\begin{figure*}[b]
    \centering
        \includegraphics[width=0.48\textwidth]{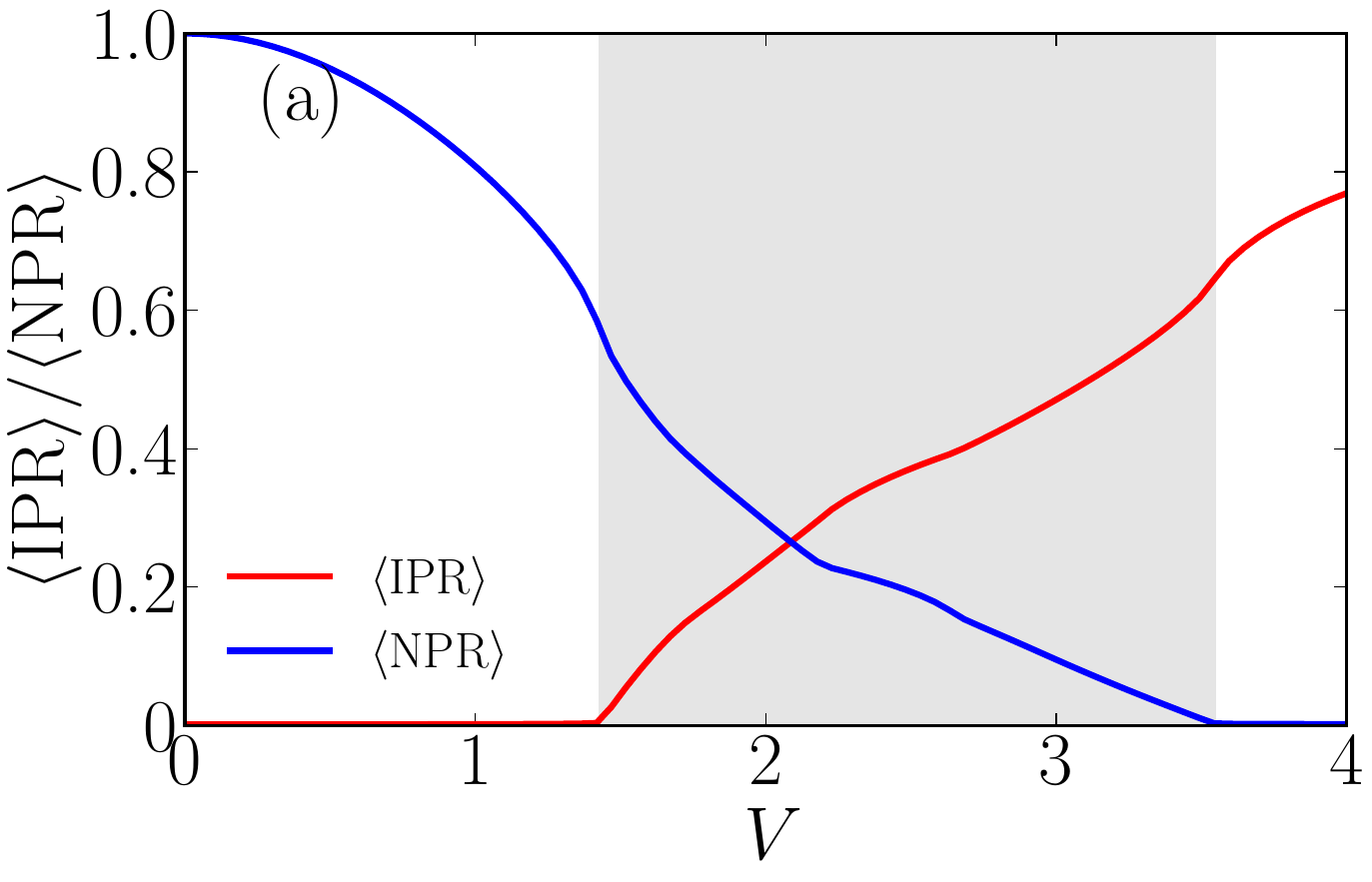}
        \includegraphics[width=0.48\textwidth]{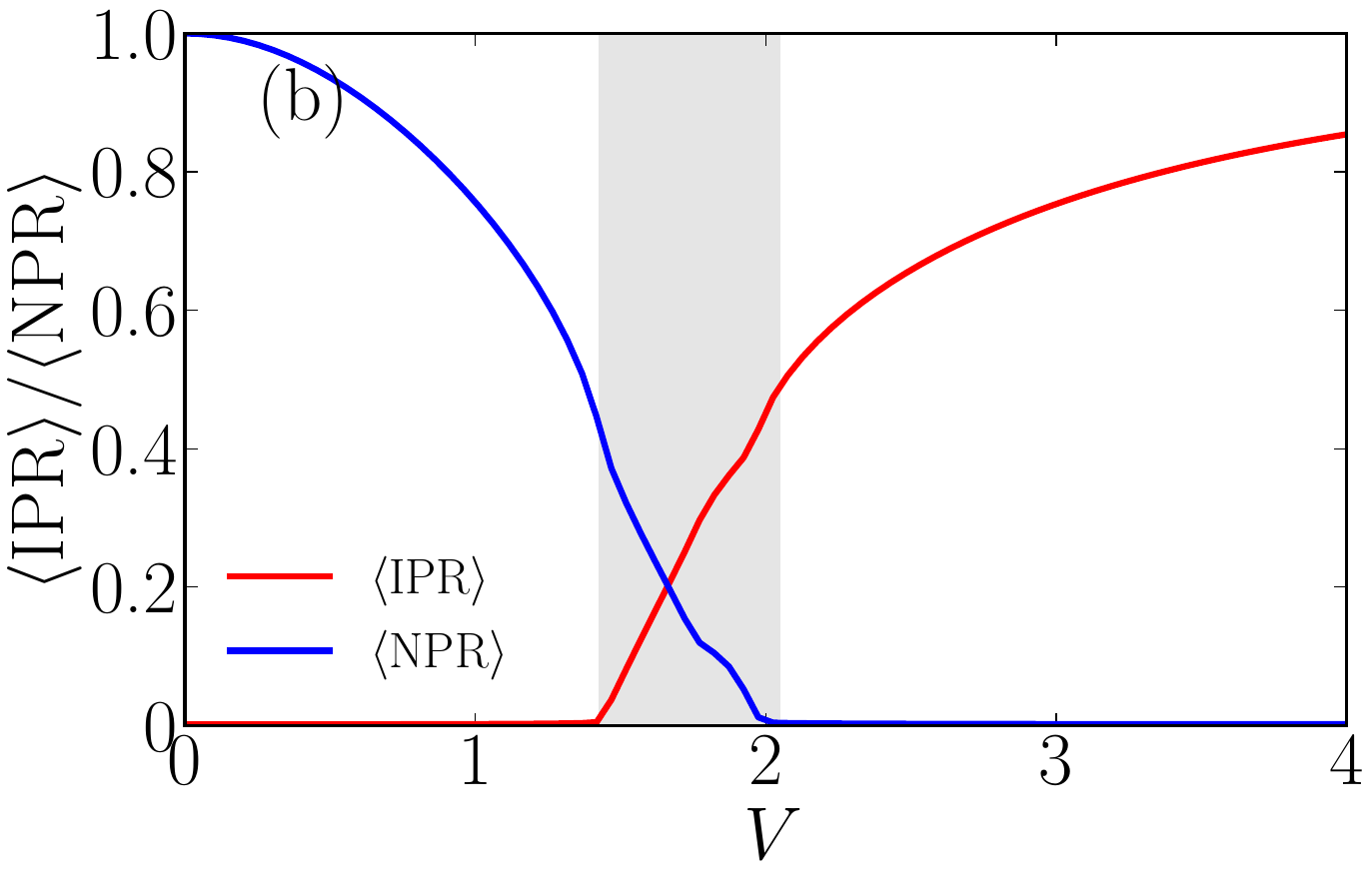}
    \caption{ Averaged IPR and NPR for all eigenstates in our model (\ref{eq1}) with (a) $\alpha_1=0.3$ and (b) $\alpha_2=0.2$, respectively. The gray regions represent the intermediate phases that hold MEs. The systems size $L=610$ and other parameters are $g\equiv h=0.5$.}
    \label{fig2}
\end{figure*}
%%%%%%%%%%%%%%%%%%%%%%%%%%%%%%%%%%%%%%%%%%%%%%%%%%%%%%%%%%%%%%%%%%%%%%%%
%%%%%%%%%%%%%%%%%%%%%%%%%%%%%%%%%%%%%%%%%%%%%%%%%%%%%%%%%%%%%%%%%%%%%%%%
\section{Numerical results}\label{section4}
We benchmark our analytical approximate ME against large-size numerical simulations using $\beta=(\sqrt{5}-1)/2$ and $L=2584$ in the PBC. To identify the extended-localized transition and characterize the extended, intermediate, and localized region, we compute the inverse participation ratio (IPR) and the  normalized participation ratio
(NPR)~\cite{qi2023multiple,li2020mobility,padhan2022emergence,li2016quantum}, which can be  written as ${\rm IPR}_n=\sum_j |\psi_j^n|^{4}$ and ${\rm NPR}_n=(L\sum_j |\psi_j^n|^{4})^{-1}$, respectively. Here, $\psi_j^n$ is the $j$-th element of the $n$-th eigenstate $|\Psi_{n}\rangle=\sum_{j=1}^{L} \psi_j^n|j\rangle$, which the Hamiltonian (\ref{eq1}) satisfies the eigenquation $H|\Psi_{n}\rangle=E_n|\Psi_{n}\rangle$. Taking the average of all the $\{{\rm IPR}_n\}$ and that of $\{{\rm NPR}_n\}$, we obtain
\begin{equation}
\langle {\rm IPR} \rangle=\frac{1}{L}\sum_{n=1}^L{{\rm IPR}_n}, \quad \langle {\rm NPR }\rangle=\frac{1}{L}\sum_{n=1}^L{{\rm NPR}_n}.
\end{equation}
Then we conclude that in the thermodynamic limit where $L$ tends to infinity,
the system is in the extended phase if $\langle{\rm IPR}\rangle \simeq 0$ and $\langle{\rm NPR}\rangle$ is finite, in the localized phase if $\langle{\rm IPR}\rangle$ is finite and $\langle{\rm NPR}\rangle \simeq 0$, and in the intermediate phase if both $\langle {\rm IPR} \rangle$ and $\langle {\rm NPR} \rangle$ are finite. As shown in Figs.~\ref{fig2} (a) and (b), at the parameters $g=h=0.5$ and the systems size $L=610$, there exist three phases as the quasiperiodic potential strength $V$ increases. For the $\alpha_1=0.3$ in the Fig.~\ref{fig2} (a), when $V<1.4$ the $\langle{\rm IPR}\rangle \simeq 0$ and $\langle{\rm NPR}\rangle$ is finite, the system is in the extended regime when $1.4\lesssim V<3.5$ the $\langle {\rm IPR} \rangle$ and $\langle {\rm NPR} \rangle$ are finite, the system is in the intermediate regime with the ME when $3.5\lesssim V$ the $\langle{\rm IPR}\rangle$ is finite and $\langle{\rm NPR}\rangle \simeq 0$, the system is in the localized regime. The similar analysis results are also found for the $\alpha_2=0.2$ and the Fig.~\ref{fig2} (b).

\begin{figure*}[t]
    \centering
    \includegraphics[width=0.98\textwidth]{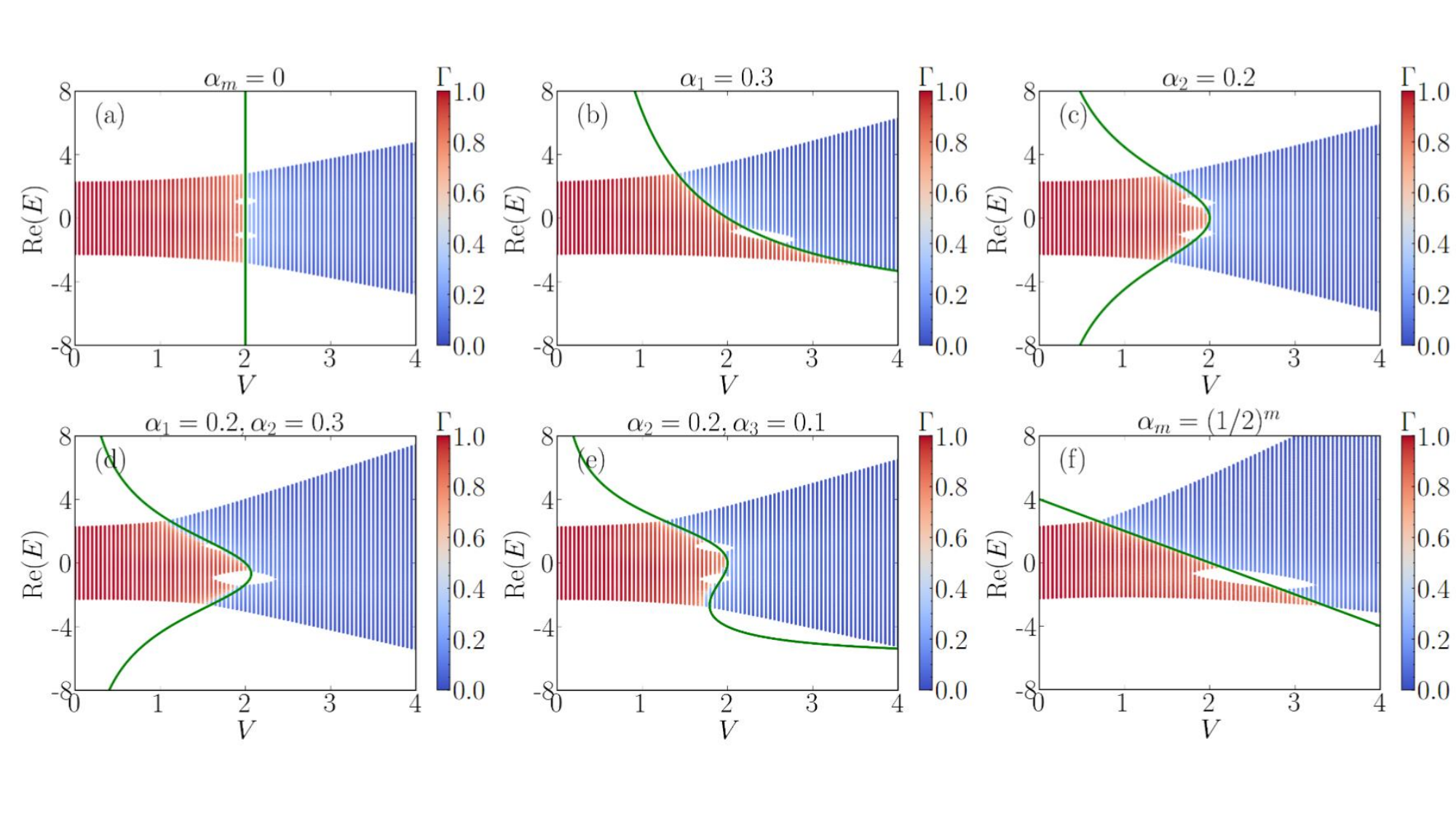}
    \vspace{-0.3in}
    \caption{Fractal dimension $\Gamma$ and non-Hermitian ME predicted from Eq.~(\ref{equation10}) (solid lines). (a) $\alpha_m=0$, (b) $\alpha_1=0.3$, (c) $\alpha_2=0.2$, (d) $\alpha_1=0.2, \alpha_2=0.3$, (e) $\alpha_2=0.2, \alpha_3=0.1$, (f) $\alpha_m=(1/2)^{m}$. The other parameters are $g\equiv h=0.5$.}
    \label{fig3}
\end{figure*}

To quantify the localization degree of each eigenstate, we introduce the fractal dimension (FD) $\Gamma_n$. For a large-size system, we can simply set $\Gamma_{n} = -\ln({\rm IPR_{n}})/\ln(L)$. Then the FD $\Gamma_n=0$ (1) represents the-$n$ maximally localized (extended) eigenstate in the thermodynamical limit, respectively. Before numerically testing the generic $\{\alpha_m\}$ case, we first discuss two limits of our ansatz, showing its agreement with the known cases of NH AAH and GPD models. The AAH model follows from Eq.~(\ref{eq2}) by putting all $\alpha_m=0$, which then gives the localization condition, according to our ansatz of Eq.~(\ref{equation10}), to be $V_c=2$. i.e., no mobility edge-- This result indicates that all states are localized (extended) for $V > V_c (V < V_c) $ and the absence of the ME. It is the connection to the GPD model is that when we set $\alpha_m=\alpha^m$, which then gives, according to our ansatz of Eq.~(\ref{equation10}), the following localization transition:
\begin{equation}
	V= 2\left[\sum_{m=0}^\infty \left(\frac{\alpha E_R}{2}\right)^m  \right]^{-1} = 2(1-\alpha E_R/2),
\end{equation}
thus the ME
\begin{equation}
	E_R = (2-V)/\alpha.
\end{equation}
This is precisely the NH GPD model analytical ME. Thus, our ansatz defined by Eq.~(\ref{equation10}) agrees with the limiting analytical results for the NH AAH and GPD models. These NH analytical ME can be confirmed in Figs.~\ref{fig3} (a) and (f), which correspond to the NH AAH and GPD models with $\alpha_m=0$ and $\alpha_m=(1/2)^m$, respectively.

To test how good our ansatz is for generic values of $\{\alpha_m\}$ for a completely general quasiperiodic potential, we show some numerical examples for a few representative situations with finite values of $\alpha_1$, $\alpha_2$, and $\alpha_3$ in Figs.~\ref{fig3} (b)-(e). It is obvious from Fig.~\ref{fig3} that our ansatz is surprisingly robust, providing $E_R$ as a function of $V$ with high accuracy. First, in Fig.~\ref{fig3} (b) we only keep nonzero $\alpha_1=0.3$ so that the system immediately has the ME as the potential broken the AAH potential in the limit $\alpha_1\to 0$.  While for $\alpha_2=0.2$, as shown in Fig.~\ref{fig3} (c), our ansatz remains a good portrayal of ME. Remarkably, even the highly nontrivial nonzero $\alpha_2$ and $\alpha_3$ structure with $E_R$ is captured correctly by our simple ansatz as shown in Figs.~\ref{fig3} (d) and (e). This indicates that the applicable range of our ME ansatz in NH quasicrystals is not simply a perturbative extension of known analytic solutions but extends far beyond. However, we also note that our ansatz is less accurate when $E_R$ approaches singularities of Eq.~(\ref{equation10}). 

\begin{figure*}[b]
    \centering
    \includegraphics[width=0.85\textwidth]{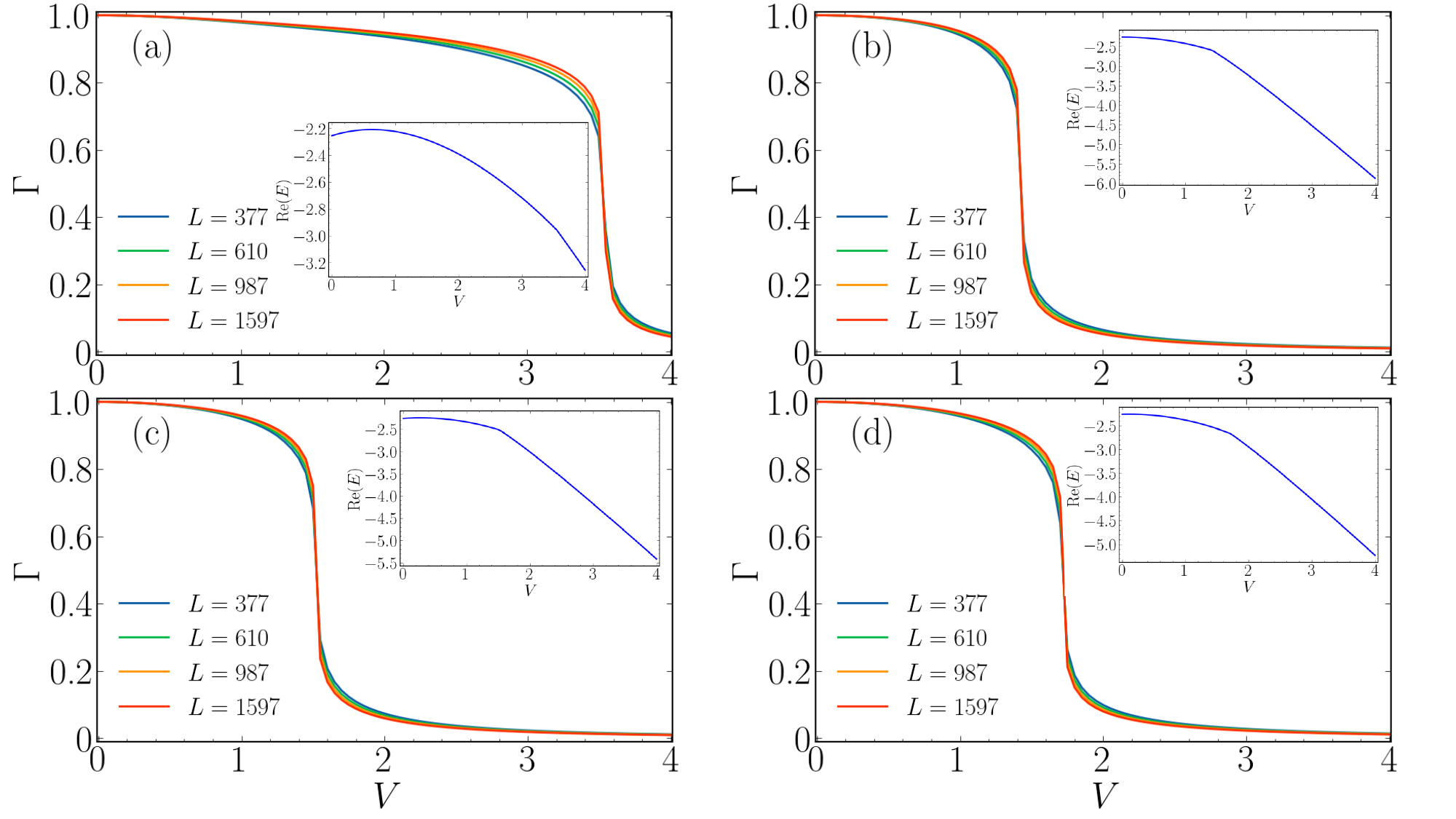}
    % \vspace{-0.3in}
    \caption{Extended-localized transition along the minimum real energy path in the $(\min(E_R), V)$ parametric phase shown in the inset obtained from finite-size scaling. (a) $\alpha_1=0.3$, (b) $\alpha_2=0.2$, (c) $\alpha_1=0.2, \alpha_2=0.3$, (d) $\alpha_2=0.2, \alpha_3=0.1$.}
    \label{fig4}
\end{figure*}
In Fig.~\ref{fig4}, we quantify the performance of our ansatz by comparing it with the critical point obtained from the standard finite-size scaling. Particularly, for $\alpha_1=0.3$ and $\alpha_2=0.3$ [Figs.~\ref{fig4} (a) and (b)], along the minimum real energy path in the $(\min(E_R), V)$ parametric phase, the ansatz Eq.~(\ref{equation10}) incurs an error of $\lesssim  5\%$. At the same time, for a more complex non-sinusoidal incommensurate potential with $\alpha_1=0.2, \alpha_2=0.3$ and $\alpha_2=0.2, \alpha_3=0.1$ [Figs.~\ref{fig4} (c) and (d)], the error is only $\lesssim  10\%$. This rather surprising finding of an approximate generic analytical solution for our NH quasiperiodic models (\ref{eq1}) hints at the complex richness of quasiperiodic localization. These approximate analytical results match the exact numerical results very well, with very small errors. This is remarkable, given that this case has no other theoretical prediction or analytical solution.

%%%%%%%%%%%%%%%%%%%%%%%%%%%%%%%%%%%%%%%%%%%%%%%%%%%%%%%%%%%%%%%%%%%%%%%%
%%%%%%%%%%%%%%%%%%%%%%%%%%%%%%%%%%%%%%%%%%%%%%%%%%%%%%%%%%%%%%%%%%%%%%%%
\section{Dynamical observation of the mobility edge}\label{section5}
\begin{figure*}[t]
    \centering
        \includegraphics[width=0.45\textwidth]{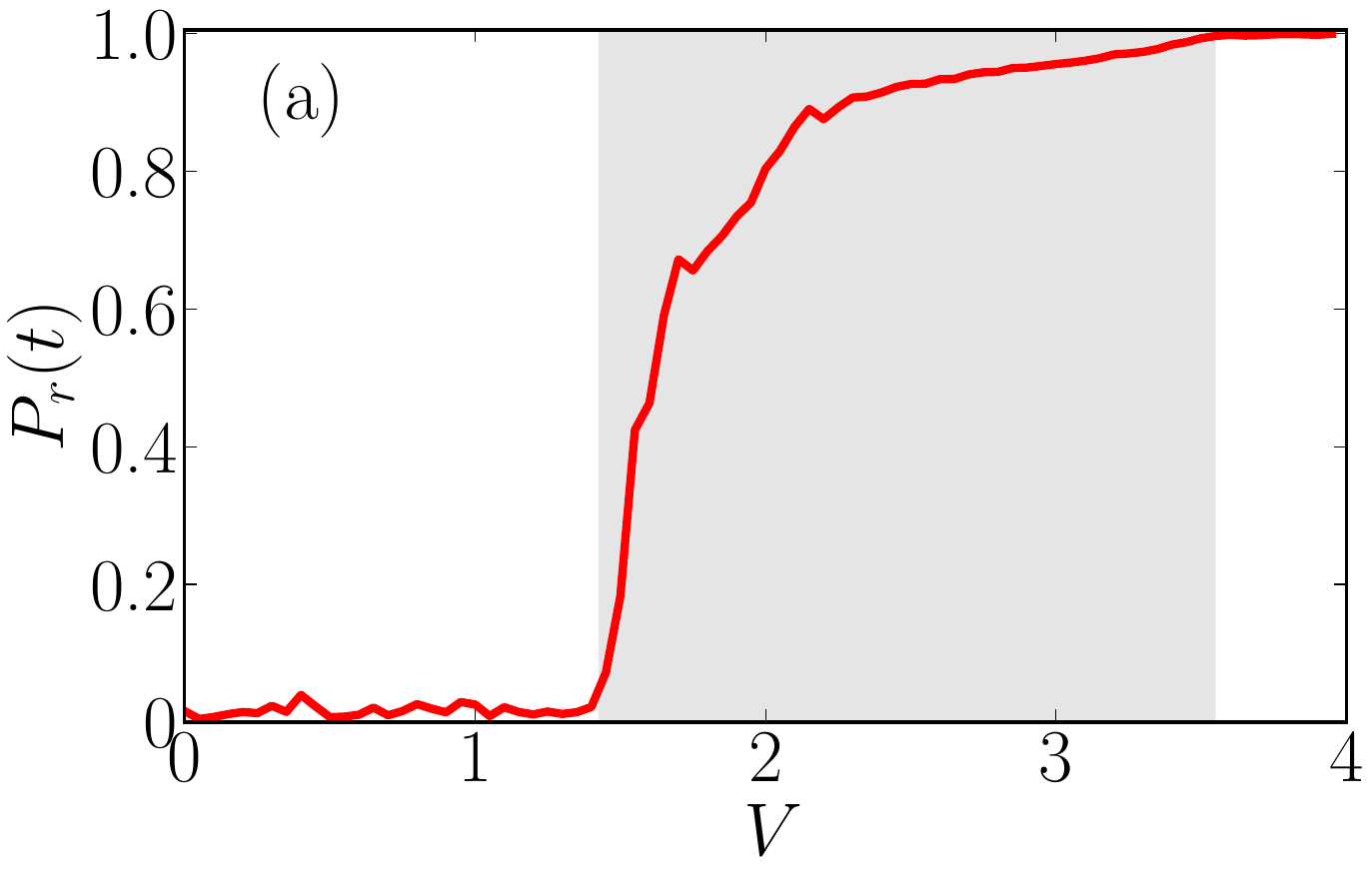}
        \includegraphics[width=0.45\textwidth]{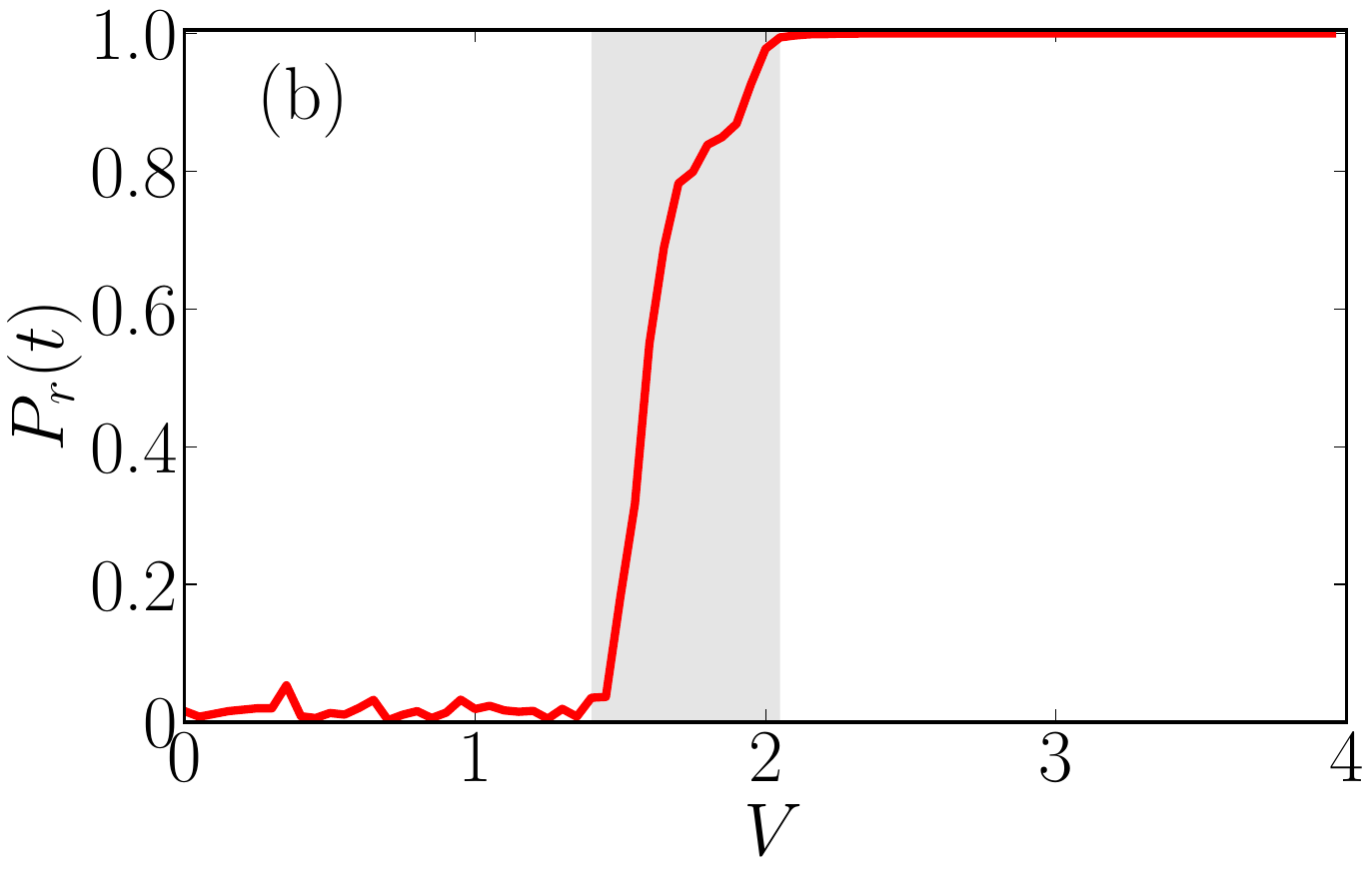}
        \\
        \includegraphics[width=0.45\textwidth]{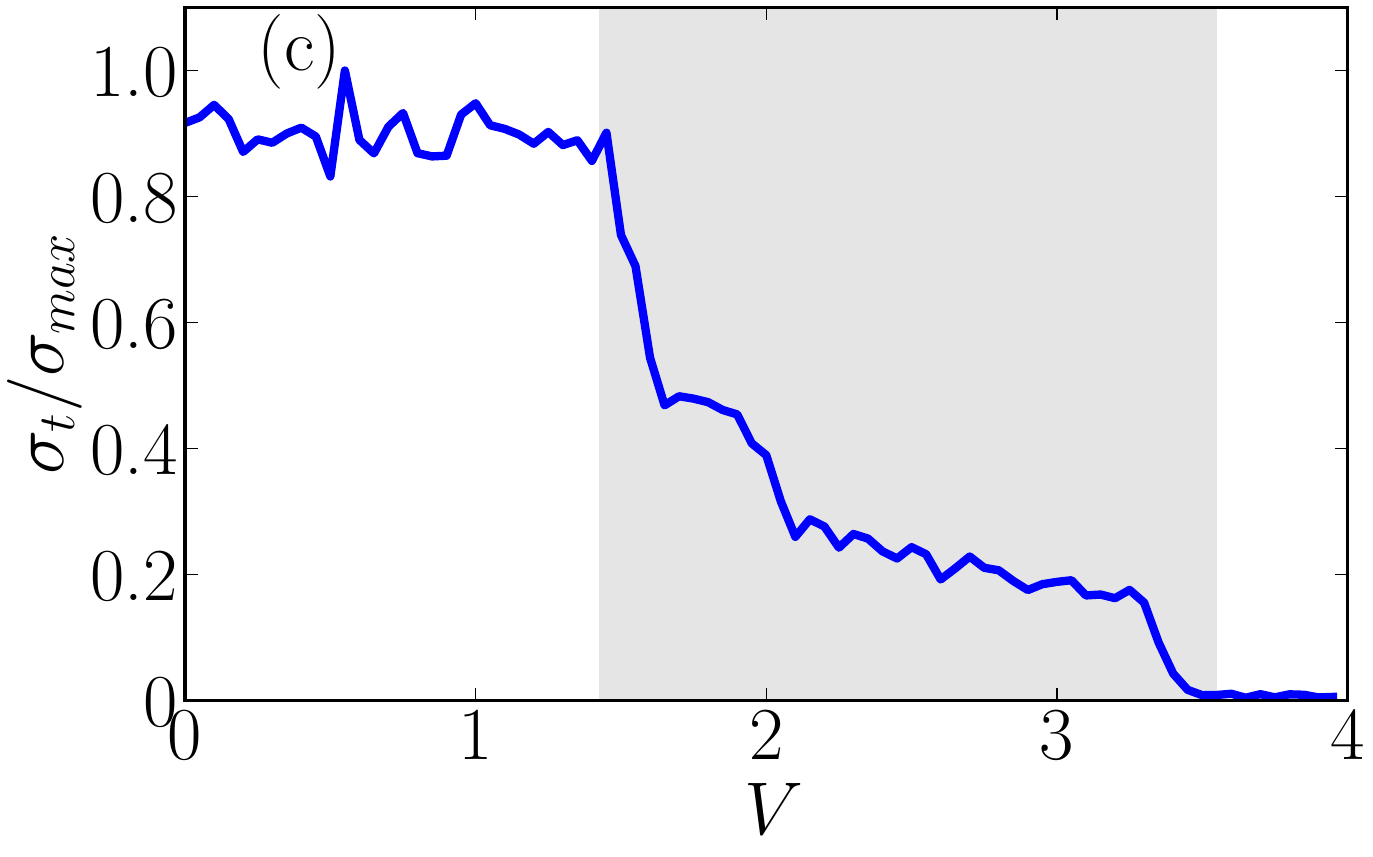}
        \includegraphics[width=0.45\textwidth]{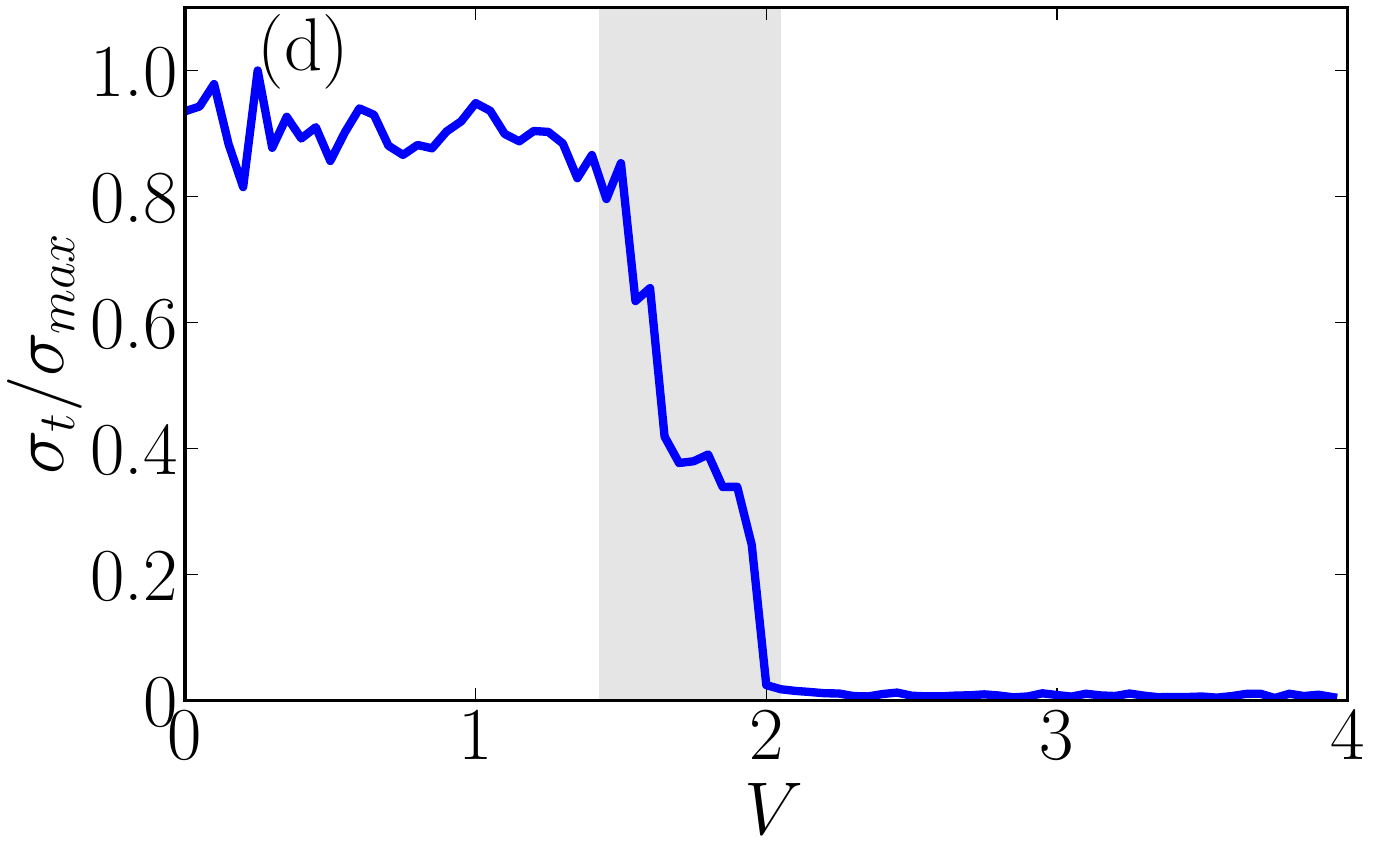}
    \caption{The survival probability $P_r(t)$ and the time evolution $\sigma_t/\sigma_{max}$ for our Hamiltonian (\ref{eq1}). (a) and (b) show the $P_r(t)$ as the function of $V$ for $\alpha_1=0.3$ and $\alpha_2=0.2$, respectively. (c) and (d) show the $\sigma_t/\sigma_{max}$ as the function of $V$ for $\alpha_1=0.3$ and $\alpha_2=0.2$, respectively. The gray regions represent the intermediate regions that hold MEs. The systems size is $L=610$, the time is $t=10^{3}$, and the other parameters are $g= h=0.1$.}
    \label{fig5}
\end{figure*}

In this section, we study the dynamic properties of the Hamiltonian (\ref{eq1}) under PBC and characterize the dynamics of different regions. Although the NH dynamics are different from the Hermitian ones, our results indicate that the dynamics and stationary states are distinct in three different regions. In an NH system, the time evolution of a given initial state $|\Psi(0)\rangle$ is determined by~\cite{hamazaki2019non,hzhai2020many,li2023non,liu2023ergodicity}
\begin{equation}
	|\Psi(t)\rangle=\frac{e^{-iHt}|\Psi(0)\rangle}{||e^{-iHt}|\Psi(0)\rangle||},
\end{equation}
where $H$ is the Hamiltonian given by Eq. (\ref{eq1}). Here the initial state is chosen as $j_0$-th basis of the Hilbert space, $|\Psi(0)\rangle=|j_0\rangle$, i.e., a single particle initially located at the $j_0$-th site of the chain at time $t=0$. Because the system (\ref{eq1}) is a 1D non-interacting tight-binding model, the involved state $|\Psi(t)\rangle$ can be decomposed as $|\Psi(t)\rangle=\sum_{j=1}^{L} \psi_j(t)|j\rangle$, where $\psi_j(t)$ is the time-dependent wave function.  By using the wave function $\psi_j(t)$, a observable physical quantity named survival probability $P_r(t)$ is proposed~\cite{deng2019one,qi2023multiple}
\begin{equation}
	P_{r}(t)=\sum_{j=\lceil \frac{L}{2} \rceil -r}^{ \lceil \frac{L}{2} \rceil +r}|\psi_j(t)|^2,
\end{equation}
where $\lceil L/2 \rceil$ means the smallest integer not less than $L/2$, and $r$ is a small integer and much smaller than the system size. This quantity describes the probability that the particle in the system's middle can still retain the initial information at a distance of $2r$ after $t$ time. Obviously, after a long evolution time, if the system is in the extended phase, the survival probability $P_{r}$ tends to 0. If the system is in the localized phase, $P_{r}$ tends to 1, while the $P_{r}$ is finite in the intermediate phase. In Figs.~\ref{fig5} (a)-(b), we insulate the numerical results of the survival probability $P_{r}$ at $t=10^{3}$ and $r=20$ and find that $P_{r}(t)$ gradually increases from $0$ to $1$ as $V$ increases. Specifically, as shown in Fig.~\ref{fig5} (a), we see that when $V<1.4$ the $P_{r}(t)$ is almost zero, the system is in the extended regime. When $3.5\lesssim V$ the $P_{r}(t)$ is almost equal to $1$, the system is in the localized regime. When $1.4\lesssim V<3.5$ the $P_{r}(t)$ is finite, and the system is in the intermediate regime holding the ME as shown in the gray region. Similar results are also shown in Fig.~\ref{fig5} (b).

Meanwhile, another dynamic quantity named mean-square displacement $\sigma(t)$ is defined as~\cite{hang2012quantum,xu2020dynamical,xu2021dynamical}
\begin{equation}\label{equation16}
	\sigma^{2}(t)=\sum_{j=1}^{L}(j-j_0)^2|\psi_j(t)|^2.
\end{equation}
Because the localized states don't diffuse in the long-time evolution, the saturation value of $\sigma(t)$ in the localized phase is smaller than that in the extended or intermediate phase.  According to Eq. (\ref{equation16}), we take $j_0 = L/2$ and denote the value of $\sigma(t)$ after a long time evolution as $\sigma_t$. We consider the rescaled quantity $\sigma_t/\sigma_{max}$, where $\sigma_{max}$ is the maximum value of $\sigma_t$ with certain model parameters in the extended phase. Then $\sigma_t/\sigma_{max}$ can distinguish the different phases. $\sigma_t/\sigma_{max}$ tends to 1 in the extended phase, tends to 0 in the localized phase, and is finite in the intermediate phase. The time evolutions of $\sigma_t/\sigma_{max}$ with the fixed parameters $g=h=0.1$ and the time $t=10^3$ are shown in Figs.~\ref{fig5} (c)-(d). We see that the $\sigma_t/\sigma_{max}$ gradually decreases from $1$ to $1$ as $V$ increases. Specifically, as shown in Fig.~\ref{fig5} (c), we see that when $V<1.4$ the $\sigma_t/\sigma_{max}$ is almost $1$, the system is in the extended regime. When $3.5\lesssim V$ the $\sigma_t/\sigma_{max}$ is almost equal to $0$, the system is in the localized regime. When $1.4\lesssim V<3.5$ the the $\sigma_t/\sigma_{max}$ is finite, and the system is in the intermediate regime holding the ME as shown in the gray region. Similar results are also shown in Fig.~\ref{fig5} (f). These results are consistent with the ones obtained by $\rm \langle IPR \rangle$ and $\rm \langle NPR \rangle$ as shown in Fig.~\ref{fig2}.

%%%%%%%%%%%%%%%%%%%%%%%%%%%%%%%%%%%%%%%%%%%%%%%%%%%%%%%%%%%%%%%%%%%%%%%%
%%%%%%%%%%%%%%%%%%%%%%%%%%%%%%%%%%%%%%%%%%%%%%%%%%%%%%%%%%%%%%%%%%%%%%%%
\section{Conclusion and outlook}\label{section6}
In conclusion, we present in this study a straightforward ansatz for the NH MEs in a class of 1D quasiperiodic models characterized by a combination of nonreciprocal hopping terms and complex quasiperiodic on-site potentials. This ansatz has been rigorously validated through exact numerical diagonalization. Our theoretical ansatz consists of two fine-tuned limiting cases that exhibit analytic MEs: the NH AAH and GPD models, and demonstrate exceptional agreement with exact numerical results across the entire parameter space for general duality-breaking quasicrystals with general values of $\{\alpha_m\}$.  The approximate ansatz yields highly accurate results that closely match the exact numerical solutions. Additionally, we explored the dynamical properties of these systems, revealing distinct behaviors in various regimes, consistent with localization transitions and MEs derived from analytical approaches. Future investigations may focus on other NH quasicrystals featuring long-range hopping terms or unbounded potentials. We also anticipate that these findings could be experimentally corroborated using NH electrical circuits, where numerous finely tuned quasiperiodic models have already been investigated~\cite{rafi2021non}. 

%%%%%%%%%%%%%%%%%%%%%%%%%%%%%%%%%%%%%%%%%%%%%%%
%%%%%%%%%%%%%%% Acknowledgement %%%%%%%%%%%%%%%
\section*{Acknowledgments}
This work is supported by the National Natural Science Foundation of China (Grant No.~62301505). LP also acknowledges support from the Fundamental Research Funds for the Central Universities.

% % \clearpage
% \appendix
% \section*{Appendix}
% \setcounter{equation}{0} \setcounter{figure}{0} \setcounter{table}{0}
% \renewcommand{\theequation}{{A}\arabic{equation}}
% \renewcommand{\thefigure}{{A}\arabic{figure}}

% \section{Non-Hermitian Ganeshan-Pixley–Das Sarma model}
% We consider the  NH Ganeshan-Pixley–Das Sarma (GPD) model, it Hamiltonian denoted as
% \begin{equation}\label{GPD}
%     H=t\sum_{j=1}^L(e^{-g}c_{j}^\dagger c_{j+1} + e^{g}c_{j+1}^\dagger c_{j}) + \sum_{j=1}^L \frac{\lambda\cos(2\pi\beta j + \theta)}{1-b\cos(2\pi\beta j + \theta)}c_j^\dagger c_j,
% \end{equation}
% where $\theta=\phi +ih $.
% By using the dual transformation
%  $u_{j,k}=\frac{1}{L}\sum_{n,m}v_{n,m}e^{-i(2\pi n\alpha_1 j+2\pi m\alpha_2 k)}$,
% Eq.~\ref{LambS} becomes
% \begin{equation}
% E v_{n,m}= V(v_{n+1,m}+v_{n-1,m}+v_{n,m+1}+v_{n,m-1})+2J[\cos(2\pi\alpha_1n)+\cos(2\pi\alpha_2m)]v_{n,m}.
% \label{LambSk}
% \end{equation}
% Eq.~\ref{LambSk} is self-dual to the original Hamiltonian defined in Eq.~\ref{LambS} when $V=J$. Thus, the extended-localized transition point is at $V=J$, and no MEs exist.

\end{document}